\pdfoutput=1
\documentclass{article}
\usepackage{spconf,amsmath,graphicx}
\usepackage{cleveref}
\Crefname{section}{Sec.}{Secs}  % https://tex.stackexchange.com/a/386310/148912

\usepackage[acronym]{glossaries}
\usepackage{balance}

\usepackage{tabularx, etoolbox, booktabs}
\usepackage{multirow} % Allows rowspan
\usepackage{subcaption}

\usepackage{pgfplots} % Import der Plots aus Matlab
\usepackage{tikz}
\pgfplotsset{compat=1.14}

\usetikzlibrary{external}
\usetikzlibrary{arrows}
\usetikzlibrary{patterns}
\usetikzlibrary{backgrounds}
\usetikzlibrary{fit}
\usetikzlibrary{positioning}
\usetikzlibrary{shapes.geometric}
\usetikzlibrary{calc}
\usetikzlibrary{shapes.multipart}
\usetikzlibrary{shapes.misc}
\usetikzlibrary{shadows.blur}

\tikzset{>=stealth}

\tikzstyle{block}=[
    draw,
    text depth=0pt, thick, rectangle, text centered,
    minimum height=3.4ex,
    fill=black!6
    ] % rectangle

\tikzstyle{branch}=[{circle,inner sep=0pt,minimum size=0.3em,fill=black}]
\tikzstyle{box}=[rectangle, rounded corners, draw=black, line width=1pt, text width=2cm]

\tikzstyle{arrow}=[{}-{>}]

\tikzset{% http://tex.stackexchange.com/a/257632
	do path picture/.style={%
		path picture={%
			\pgfpointdiff{\pgfpointanchor{path picture bounding box}{south west}}%
			{\pgfpointanchor{path picture bounding box}{north east}}%
			\pgfgetlastxy\x\y%
			\tikzset{x=\x/2,y=\y/2}%
			#1
		}
	},
	sin wave/.style={do path picture={    
			\draw [line cap=round] (-3/4,0)
			sin (-3/8,1/2) cos (0,0) sin (3/8,-1/2) cos (3/4,0);
	}},
	cross/.style={do path picture={    
			\draw [line cap=round] (-2/5,-2/5) -- (2/5,2/5) (-2/5,2/5) -- (2/5,-2/5);
	}},
	plus/.style={draw, circle, do path picture={    
			\draw [line cap=round] (-3/5,0) -- (3/5,0) (0,-3/5) -- (0,3/5);
	}},
	mic/.style={inner sep=0pt, do path picture={
			\draw (0,0) circle (0.9);
			\draw [line cap=round] (-0.9, -0.9) -- (-0.9, 0.9);
	}},
	mux/.style={trapezium, draw}
}

\usepackage{amsmath}
\usepackage{amssymb}
\usepackage{bbm}
\usepackage{physics}
\usepackage{siunitx}
\usepackage{trfsigns}

\DeclareMathOperator{\ExpOp}{\mathbb{E}}

\newcommand{\vect}[1]{\ensuremath{\boldsymbol{\mathbf{#1}}}}

\renewcommand{\H}{^\mathrm{H}}
\newcommand{\T}{^\mathrm{T}}
\newcommand{\inv}{^{-1}}

\def\x{{\mathbf x}}

\newacronym{ASR}{ASR}{Automatic Speech Recognition}
\newacronym{ATF}{ATF}{acoustic transfer function}
\newacronym{BF}{BF}{beamforming}
\newacronym{CNN}{CNN}{Convolutional Neural Network}
\newacronym{CTF}{CTF}{convolutive transfer function}
\newacronym{EM}{EM}{Expectation Maximization}
\newacronym{GSC}{GSC}{Generalized Sidelobe Canceller}
\newacronym{GMM}{GMM}{Gaussian Mixture Model}
\newacronym{GSS}{GSS}{Guided Source Separation}
\newacronym{LF-MMI}{LF-MMI}{Lattice-Free Maximum Mutual Information}
\newacronym{MIMO}{MIMO}{Multiple Input Multiple Output}
\newacronym{ML}{ML}{maximum likelihood}
\newacronym{MLDR}{MLDR}{Maximum Likelihood Distortionless Response}
\newacronym{MM}{MM}{Mixture Model}
\newacronym{MPDR}{MPDR}{Minimum-Power Distortionless Response}
\newacronym{MVDR}{MVDR}{Minimum Variance Distortionless Response}
\newacronym{NN}{NN}{neural network}
\newacronym{PSD}{PSD}{Power Spectral Density}
\newacronym{RIR}{RIR}{room impulse response}
\newacronym{RTF}{RTF}{relative transfer function}
\newacronym{TDNN}{TDNN}{Time Delay Neural Network}
\newacronym{TDNN-F}{TDNN-F}{factorized TDNN}
\newacronym{STFT}{STFT}{Short Time Fourier Transform}
\newacronym{WER}{WER}{word error rate}
\newacronym{wMPDR}{wMPDR}{weighted MPDR}
\newacronym{WPE}{WPE}{Weighted Prediction Error}
\newacronym{WPD}{WPD}{weighted power minimization distortionless response}

\title{Jointly optimal dereverberation and beamforming}
\name{Christoph Boeddeker$^{1}$, Tomohiro Nakatani$^{2}$, Keisuke Kinoshita$^{2}$, Reinhold Haeb-Umbach$^{1}$}
\address{$^{1}$ Paderborn University, Department of Communications Engineering, Paderborn, Germany \\
$^{2}$ NTT Communication Science Laboratories, NTT Corporation, Kyoto, Japan
}

\begin{document}
\ninept
\maketitle
% http://thomas.deselaers.de/computing/texsqueezing.html
% \the\abovedisplayskip  % see the actual value
% \the\belowdisplayskip  % see the actual value
% \abovedisplayskip8.5pt plus 3.0pt minus 4.0pt  % default
% \belowdisplayskip8.5pt plus 3.0pt minus 4.0pt  % default
\abovedisplayskip7.pt plus 3.0pt minus 4.0pt
\belowdisplayskip7.pt plus 3.0pt minus 4.0pt
\setlength{\textfloatsep}{1em}
\begin{abstract}
We previously proposed an optimal (in the maximum likelihood sense) 
convolutional beamformer that can perform simultaneous denoising and 
dereverberation, and showed its superiority over the widely used cascade  of a \gls{WPE} dereverberation filter and a 
conventional \gls{MPDR} beamformer. 
However, it has not been fully investigated which components in the 
convolutional beamformer yield such superiority. To this end, this paper 
presents a new derivation of the convolutional beamformer that allows us 
to factorize it into a \gls{WPE} dereverberation filter, and a 
special type of a (non-convolutional) beamformer, referred to as a 
\gls{wMPDR} beamformer, without loss of optimality. With 
experiments, we show that the superiority of the convolutional 
beamformer in fact comes from its \gls{wMPDR} part.
% We present two derivations of an optimal (in the Maximum Likelihood sense) convolutional beamformer for simultaneous dereverberation and denoising.
% The first assumes that the estimator consists of a single linear filter, while the second assumes a MIMO filter matrix followed by a spatial filter, according the common processing flow of dereverberation followed by beamforming.
% While both approaches have been presented in the literature, we here show that they are equivalent, leading to the exactly same output signal.
% Experimental results demonstrate the effectiveness of both dereverberation and beamforming in terms of error rate improvements of a downstream speech recognizer.
\end{abstract}
\glsreset{wMPDR}
\section{Introduction}
\label{sec:intro}
In many speech processing applications the microphone signal is degraded both by reverberation and by noise. 
Reverberation is caused by the  signal traveling from source to the sensor via multiple paths with different lengths and attenuations, causing a temporal smearing.
While early reflections which arrive with up to roughly \SI{50}{ms} delay compared to the direct signal are actually beneficial for human perception and even for \gls{ASR}, late reverberation degrades both of them.
Furthermore, if microphones are located at a distance to the speaker, it is likely that they capture other signals, here denoted as noise for simplicity, in addition to the desired speech signal.

For improving the quality of recorded speech, a number of techniques has been developed for dereverberation and denoising.
Dereverberation techniques can be broadly categorized \cite{LiDeHaGo2015} into spectral magnitude manipulation \cite{Lebart2001AA} and linear filtering techniques \cite{Schwarts2015TASLP,Nakatani2010TASLP}.
Among the latter, the \gls{WPE} method \cite{Nakatani2010TASLP,Yoshioka2012GeneralWPE}, has been shown to be very effective both for improving signal quality for human perception and for ASR \cite{REVERB2016}.
A very effective mean to remove noise is acoustic beamforming based on microphone arrays \cite{VanVeen1988Beamforming,Erdogan2015masks}, which can enhance the desired source signal while attenuating signals with different propagation patterns.
%Beamforming was also successfully applied to dereverberation  \cite{Schwarts2015TASLP}.
Furthermore, for performing dereverberation and denoising at the same time, their cascade configuration has been widely studied.
Its effectiveness was shown by top scoring systems developed for recent distant speech recognition challenges, such as the REVERB and the CHiME-3/4/5 challenges \cite{REVERB2016,Barker2015CHiME3,Kanda2019}.
Iterative optimization of the cascade configuration is also investigated as extension of this approach \cite{Braun2018TASLP,Drude2018integration}.

One issue of the conventional cascade approach \cite{Drude2018integration,Delcroix2015ESP} was that the overall optimality was not guaranteed.
The approach performs the optimization separately for dereverberation and beamforming. Moreover, different optimization criteria are adopted for the respective problems:
The \gls{WPE} technique estimates the dereverberation filter based on maximum likelihood estimation with a time-varying Gaussian source assumption \cite{Nakatani2010TASLP}, while most beamformers are estimated based on a noise power minimization criterion \cite{VanVeen1988Beamforming}.
As a consequence, what is optimized by the cascaded approach was even not clear.
Also, it was not well investigated what optimization criterion is preferable for simultaneous denoising and dereverberation.

% \inred{An first investigation of integration of \gls{WPE} and beamforming was done in \cite{Drude2018integration}. While the derivation of \gls{WPE} considered the beamformer
% using iterative optimization is also
% investegated as extension of this approach\cite{Drude2018integration}.}

To address this issue, a convolutional beamformer has been recently proposed \cite{Nakatani2019AUC}.
It unifies the WPE dereverberation filter and a beamformer into a single linear convolutional filter, called \gls{WPD} beamformer.
The filter coefficients are optimized based on a single criterion, namely the likelihood maximization with a time-varying Gaussian source assumption \cite{nakatani2019maximum}.
The \gls{WPD} beamformer was compared with a conventional cascade configuration, consisting of a \gls{WPE} \gls{MIMO} dereverberation filter \cite{Yoshioka2012GeneralWPE} followed by an \gls{MPDR} beamformer. 
Experiments carried out on the REVERB challenge data set showed a performance advantage in terms of \gls{WER} of the \gls{WPD} over the cascade structure.

However, it remains unclear what makes the WPD beamformer superior to the conventional cascade configuration, and if at all there is an essential difference between the unified and cascade structures.
%how and if at all there is an essential difference between the unified and cascade structures. 
This paper answers these questions and shows the equivalence between the \gls{WPD} convolutional beamforming and the cascade configuration of \gls{MIMO}-\gls{WPE} dereverberation followed by a variant of MPDR beamforming, namely \gls{wMPDR} beamforming.
We theoretically derive their strict equivalence under the assumption that the two  are jointly optimized based on the maximum likelihood criterion.   
%and shows theoretically and experimentally that the  \gls{WPD} and the cascade of MIMO-WPE followed by a variant of MPDR, namely a \gls{wMPDR}, deliver the identical output signal when the cascaded processing blocks are jointly optimized based on the same maximum-likelihood criterion. 
%Thus, we can factorize WPD convolutional beamforming into WPE MIMO dereverberation and \gls{wMPDR} beamforming without loss of optimality.
%In experiments, we further 
%Then, by experiments, we show that even when each processing is separately optimized the cascade configuration can achieve comparable performance with the \gls{WPD} beamforming. 
%
The factorizability of the convolutional beamforming has some practical advantages due to its modularity.
For example, signal parameters, such as the spatial covariance matrix of the vector of microphone signals and the time variant clean speech power, need to be estimated from the data.
In \cite{nakatani2019maximum}, this was done with the help of additional WPE preprocessing.
With the result given here, this WPE component can be a part of the enhancement chain, thus simplifying the overall structure.

We further show that the performance advantage obtained by the WPD beamforming or its equivalent cascaded structure over the conventional cascade structure comes from the \gls{wMPDR} beamforming over conventional beamforming.

The paper is organized as follows:
After the signal model is introduced in \Cref{sec:signalmodel}, unified and factorized versions of the convolutional beamformer are derived in Secs. \ref{sec:unified} and \ref{sec:cascaded}.
\Cref{sec:discussion} discusses the characteristics of the factorized and unifed structure referring to its equivalence shown in the Appendix.
Experimental validation of the  theory and concluding remarks are given in Secs.~\ref{sec:experiments} and \ref{sec:conclusions}.

\newcommand{\shat}{\hat{s}}

\renewcommand{\x}{\vect{x}}
\newcommand{\y}{\vect{y}}
\newcommand{\w}{\vect{w}}
\renewcommand{\d}{\vect{d}}
\newcommand{\dhat}{\breve{\vect{d}}}
\newcommand{\wpast}{\tilde{\vect{w}}}
\newcommand{\wstacked}{\bar{\vect{w}}}
\renewcommand{\a}{\vect{a}}
\newcommand{\R}{\vect{R}_{\y}}
\newcommand{\Rd}{\vect{R}_{\d}}
\newcommand{\G}{\vect{G}}
\newcommand{\Gpast}{\tilde{\vect{G}}}
\newcommand{\Gstacked}{\bar{\vect{G}}}
\newcommand{\Rpast}{\tilde{\vect{R}}_{\y}}
\renewcommand{\P}{\vect{P}_{\y}}
\newcommand{\xstacked}{\bar{\vect{x}}}
\newcommand{\ypast}{\tilde{\vect{y}}}
\newcommand{\ystacked}{\bar{\vect{y}}}
\newcommand{\apure}{\vect{v}}
\newcommand{\rtf}{\tilde{\vect{v}}}
\newcommand{\astacked}{\bar{\vect{v}}}
\newcommand{\Rstacked}{\bar{\vect{R}}_{\y}}

\newcommand{\bfonly}{\vect{q}}
\newcommand{\bfonlystacked}{\bar{\vect{q}}}

\section{Signal Model}
\label{sec:signalmodel}
We assume that a single speech signal is captured by $M$ microphones in a noisy and reverberant environment.
In the \gls{STFT} domain the vector of microphone signals $\y_{t} = \begin{bmatrix} y_{1,t} &\ldots & y_{M,t}\end{bmatrix}\T$ can be written as the convolution of the source signal $s_t$ with the vector of the convolutive transfer function $\a_{\tau}=\begin{bmatrix} a_{1,\tau} &\ldots & a_{M,\tau}\end{bmatrix}\T$ plus an additive noise vector $\vect{n}_t$:
\begin{align}
	\y_{t} &= \sum_{\tau=0}^{L_a-1}\a_{\tau}s_{t-\tau} + \vect{n}_{t}\label{eq:signal_model}\\
	 &=\vect{x}_{t} + \vect{n}_{t} 
	 = \vect{d}_{t} + \vect{r}_{t} +\vect{n}_{t}.
\end{align}
Here, $t$ is the time frame index.
The frequency bin index has been dropped for ease of notation.
$L_a$ is the length of the transfer function in number of frames. The term $\vect{x}_t$ is called the image of the source signal $s_t$ at the microphones, which is further decomposed in the direct signal plus early reflections $\vect{d}_{t}$, and late reverberation $\vect{r}_{t}$:
\begin{align}
  \label{eq:desired_signal_approx}
	\vect{d}_{t} &= \sum_{\tau=0}^{b-1} \a_{\tau}s_{t-\tau} \approx \apure s_t = \rtf d_{1,t} \\
	\vect{r}_{t} &= \sum_{\tau=b}^{L_a-1} \a_{\tau}s_{t-\tau},
\end{align}
where the frame index $b$ separates the early reflections from the late reverberation. A typical value for $b$ is 2 to 4 frames, corresponding to $\num{30}$ to $\SI{50}{ms}$.
In \eqref{eq:desired_signal_approx} we approximated $\vect{d}_{t}$ by the product of a time-invariant (non-convolutive) \gls{ATF} vector $\vect{v}$ with the clean speech signal $s_t$. Furthermore we introduced the \gls{RTF} $\rtf= \apure/v_1$, and $d_{1,t}= v_1s_t$. 

We now define the vector of the past $L_w$ microphone signals, including the current observation $\y_t$, but excluding the most recent $b-1$ frames:
\begin{align}
	\ystacked_t &= \begin{bmatrix}	\y_{t}\T &  \y_{t-b}\T & \ldots & 	\y_{t-L_w+1}\T 	\end{bmatrix}\T \in \mathbb{C}^{M(L_w-b+1)\times 1}\\
	 &= \begin{bmatrix} \y_{t}\T  & \ypast_{t}\T  \end{bmatrix}\T
\end{align}
where $\ypast$ captures the observations from $b$ frames in the past until $L_w-1$ frames in the past.
Our goal is to determine the coefficients $\wstacked$ of a spatial filter such that 
\begin{align}
	z_t = \wstacked\H \ystacked_t \label{eg:estimate}
\end{align}
is an estimate of the desired signal $d_{1,t}$. Here, $\wstacked$ is the vector
\begin{align}
	\wstacked &= \begin{bmatrix}	\w_{0}\T & \w_b\T & \ldots &	\w_{L_w-1}\T	\end{bmatrix}\T,
\end{align}
which has the same dimension as $\ystacked$.
Because this beamformer is based on the convolutional signal model of eq.~\eqref{eq:signal_model} we call it \textit{convolutional beamformer} \cite{Nakatani2019AUC}.

%%%%%%%%%%%%%%%%%%%%%%%%%%%%%%%%%%%%%%%%%%%%%%%%%%%%%%%%%%%%%%%%%%%%%
\section{Unified solution}
\label{sec:unified}
In \cite{nakatani2019maximum}, the output $z_t$ is modeled as a zero mean complex Gaussian with a time varying variance.
This output distribution was used to define the \gls{ML} objective for the estimation of the coefficients $\wstacked$.
Under a distortionless response constraint that is often introduced into beamforming the \gls{ML} objective can be replaced with: 
\begin{align}
	\mathcal{L}(\wstacked) &\propto \frac{1}{T} \sum_{t=1}^{T} \left(- \ln(\lambda_t) -  \frac{\left|z_t\right|^2}{\lambda_t} \right)\label{eq:likelihood}\\
	&\propto - \wstacked\H  \left(\frac{1}{T}\sum_{t=1}^{T} \frac{\ystacked_t\ystacked_t\H}{\lambda_t}\right)\wstacked
	= - \wstacked\H\Rstacked\wstacked, \label{objective_1}
\end{align}
where $\lambda_t = \ExpOp[\left|d_{1,t} \right|^2]$, $\Rstacked = \frac{1}{T} \sum_t \ystacked_t\ystacked_t\H / \lambda_t$, and where $T$ is the number of frames over which the beamformer coefficients are estimated.
% Optimizing this function w.r.t. $\wstacked$ obviously needs a constraint.
% A natural choice is to postulate a distortionless response:
The distortionless response constraint introduced for the \gls{ML} estimation was:
\begin{align}
	\w_0\H\rtf = 1.
\end{align}
This can be reformulated by introducing $\wstacked$ as follows:
\begin{align}
  \label{eq:def_wstacked}
	\wstacked\H\astacked = 1,
\end{align}
where $\astacked = \begin{bmatrix} \apure\T/v_1 & \vect{0}\T \end{bmatrix}\T$, and where $\vect{0}$ is a vector of zeros of  dimension $(M\cdot(L_w-b)\times 1)$. 
A constrained optimization problem
\begin{align}
  \label{sec:optimal}
	\mathcal{L}(\wstacked) &= -\wstacked\H\Rstacked\wstacked \quad\text{s.t.}\quad\wstacked\H\astacked = 1
\end{align}
of this kind  is well-known from minimum variance/power distortionless beamforming, and the solution is given by
\begin{align}
	\wstacked &= \frac{\Rstacked\inv\astacked}{\astacked\H\Rstacked\inv\astacked}.\label{eq:wpdfilter}
\end{align}
This is the \gls{WPD} beamformer proposed in \cite{Nakatani2019AUC}.

%%%%%%%%%%%%%%%%%%%%%%%%%%%%%%%%%%%%%%%%%%%%%%%%%%%%%%%%%%%%%%%%%%%%%
\section{Factorized solution}
\label{sec:cascaded}
%%%%%%%%%%%%%%%%%%%%%%%%%%%%%%%%%%%%%%%%%%%%%%%%%%%%%%%%%%%%%%%%%%%%%

Now we assume that $\wstacked$ factorizes into a $(M(L_w-b+1) \times M)$-dimensional \gls{MIMO} dereverberation matrix $\Gstacked$ and a beamforming vector $\bfonly$ of size $(M\times 1)$
\begin{align}
  \label{eq:factorized_1}
	\wstacked = \Gstacked\bfonly.
\end{align}
Using this in the objective function~\eqref{objective_1}  we obtain
\begin{align}
	\label{greedyLikelihood} 
	\mathcal{L}(\Gstacked,\bfonly) &\propto -\bfonly\H\Gstacked\H \Rstacked \Gstacked\bfonly. 
\end{align}
Note that $\Gstacked$ has a particular structure, because we have to make sure that the direct signal and early reflections remain unmodified by the derevereberation matrix: 
\begin{align}
  \label{eq:Gstacked}
	\Gstacked = \begin{bmatrix} \vect{I}_M \\ -\Gpast\end{bmatrix}.
\end{align}
Here, $\vect{I}_M$ is the $(M\times M)$-dimensional identity matrix, and $\Gpast$ is of dimension $(M(L_w-b)\times M)$.
Similarly, we factorize $\Rstacked$:
\begin{align}
  \label{eq:Rstacked}
	\Rstacked = \begin{bmatrix} \R & \P\H \\ \P & \Rpast \end{bmatrix}
\end{align}
where	$\R = \frac{1}{T} \sum_{t=1}^{T} \y_t\y_t\H/\lambda_t$,
	$\P = \frac{1}{T} \sum_{t=1}^{T} \ypast_t\y_t\H /\lambda_t$, and $\Rpast = \frac{1}{T} \sum_{t=1}^{T} \ypast_t\ypast_t\H/\lambda_t$.
Using this in \eqref{greedyLikelihood}, we arive at
\begin{align}
	&\mathcal{L}(\Gpast,\bfonly) \propto -\bfonly\H\begin{bmatrix}		\vect{I} \\ -\Gpast \end{bmatrix}\H
	\begin{bmatrix} 	\R & \P\H \\ \P & \Rpast 	\end{bmatrix}
	\begin{bmatrix} \vect{I} \\ -\Gpast 	\end{bmatrix}\bfonly \nonumber \\
	&= -\bfonly \R\bfonly + \bfonly\H\Gpast\H\P\bfonly + \bfonly\H\P\H\Gpast\bfonly - \bfonly\H\Gpast\H\Rpast\Gpast\bfonly.
	\label{optimize_2}
\end{align}
To calculate the derivative w.r.t. the dereverberation matrix we use eq. (70), (71) and (82) from \cite{Petersen2008MCB} and the property $\Rpast = \Rpast\H$. Setting the derivative to zero gives
\begin{align}
	\pdv{\mathcal{L}(\Gpast,\bfonly)}{\Gpast} 
	&= 2\P\bfonly\bfonly\H - 2 \Rpast\Gpast\bfonly\bfonly\H \stackrel{!}{=} \vect{0}.
\end{align}
This equation has obviously multiple solutions.
%So there is no unique solution.
A solution, which allows separate estimation of $\Gpast$ and $\bfonly$ is
\begin{align}
  \label{eq:Gpast}
	\Gpast &= \Rpast\inv\P.
\end{align}
This solution is identical to the WPE solution \cite{Yoshioka2012GeneralWPE,Drude2018narawpe,Boeddeker2018MA} (Scaled Identity Matrix assumption: $\dhat \sim \mathcal{N}(\vect{0}, \lambda_t \vect{I})$). It allows to  obtain a first estimate the dereverberated signal from the current observation via
\begin{align}
	\dhat_t = \Gstacked\H\ystacked_t.
\end{align}
%the late reverberation present in the current observation
%\begin{align}
%	\hat{\vect{r}}_t = \Gpast\H\ypast_t.
%\end{align}
Next we optimize \eqref{greedyLikelihood} w.r.t. the beamforming vector. This is analog to \cref{sec:optimal}, if $\wstacked$ and $\Rstacked$ are replaced  by $\bfonly$ and $\Rd = \Gstacked\H\Rstacked\Gstacked$, respectively. Thus, using the constraint $\bfonly\H\rtf=1$ gives the solution:
\begin{align}
  \label{eq:mvdr_like}
	\bfonly &= \frac{\Rd\inv\rtf}{\rtf\H\Rd\inv\rtf}.
\end{align}
$\Rd$ can be estimated just like we estimated $\Rstacked$ above:
\begin{align}
  \label{eq:Rd}
	\Rd	
	= \frac{1}{T} \sum_{t} \Gstacked\H\frac{\ystacked_t\ystacked_t\H}{\lambda_t}\Gstacked
	= \frac{1}{T} \sum_{t} \frac{\dhat_t\dhat_t\H}{\lambda_t}.
	\end{align}
%where $\hat{\vect{d}} = \Gstacked\H\ystacked$.\glsreset{wMPDR}
This beamformer is similar to the \gls{MPDR} beamformer, but the denominator in eq.~\eqref{eq:Rd} makes this beamformer a \gls{wMPDR}.
%, also known as a \gls{MLDR} beamformer \cite{Cho2019MLDR}. 
It is worth noting that this beamformer can be derived as a special case of WPD by assuming the absence of reverberation and setting the length of the convolutional beamformer in (\ref{eq:signal_model})-(\ref{eq:wpdfilter}) to $L_w=1$.
This beamformer was independently proposed as a \gls{MLDR} in \cite{Cho2019MLDR}.
%Similar to the motivation of the W in WPE [Nakatani???], we call this beamformer weighted MPDR (wMPDR).
%\inred{Hier auf Park verweisen?}

% eg:estimate
 
%%%%%%%%%%%%%%%%%%%%%%%%%%%%%%%%%%%%%%%%%%%%%%%%%%%%%%%%
\section{Discussion}
\label{sec:discussion}
%%%%%%%%%%%%%%%%%%%%%%%%%%%%%%%%%%%%%%%%%%%%%%%%%%%%%%%%

In the appendix  we show that the unified (\gls{WPD}) beamformer and factorized solution, which consists of the cascade of \gls{WPE} dereverberation and \gls{wMPDR} beamforming, are identical.
Instead of estimating the coefficient vector of the convolutional beamformer, it is thus equivalent to first dereverberate the vector of microphone signals using the MIMO-WPE method and then applying a wMPDR beamformer resting on the narrowband assumption to the result.
This cascaded solution may have some practical advantages, because it allows to treat dereverberation and beamforming separately.
Although the equivalence has only been derived for the ML convolutional beamformer, it may still be seen as an indication, that a cascade of dereverberation with a beamformer optimized under another criterion is a legitimate solution as done in \cite{Drude2018integration}.
%It is also worthwhile noting that in the absence of reverberation the derived beamformer simplifies into the solution proposed in \cite{Cho2019MLDR}. 

Comparing the constrained optimization problem \eqref{sec:optimal} with the classical \gls{MPDR}, the difference is the scaling of the beamformer output power by the variance of the clean speech signal.
This scaling in the objective function accounts for the time-varying nature of the speech power.
Observations with large speech variance are downscaled, while observations with a low variance are emphasized for spatial covariance estimation for beamforming coefficient computation.
This makes sense because we do not want to destroy the speech signal and only suppress the distortions. 
 
In practice the parameters of the statistical models involved have to be estimated from the data.
This includes the \gls{RTF} $\rtf$ and the time-variant power spectral density of the desired speech component $\lambda_t$, which will be discussed in the next section.
%, and the spatial covariance matrices $\Rstacked$.

%%%%%%%%%%%%%%%%%%%%%%%%%%%%%%%%%%%%%%%%%%%%%%%%%%%%%%%%
\section{Experiments}
\label{sec:experiments}
%%%%%%%%%%%%%%%%%%%%%%%%%%%%%%%%%%%%%%%%%%%%%%%%%%%%%%%%

%\inblue{Comment how to best estimate params $\lambda_t$ and spatial covariance matrix, and \gls{RTF}.
%Include similar block diagram as in EUSIPCO paper.
%Discuss figure of evaluation system.
%
%Demonstrate importance of $\lambda_t$ with ASR results.

\begin{figure}
	\centering
  \tikzset{pics/.cd,
 	pic switch/.style args={#1 times #2}{code={
 			% \node[draw, fit={#1#2}] {box};
 			\tikzset{x=#1/2,y=#2/2}
 			\coordinate (-north west) at (-1,1);
 			\coordinate (-north east) at (1,1);
 			\coordinate (-south west) at (-1,-1);
 			\coordinate (-south east) at (1,-1);
 			\coordinate (-north) at (0,1);
 			\coordinate (-east) at (1,0);
 			\coordinate (-south) at (0,-1);
 			\coordinate (-west) at (-1,0);
 			\coordinate (-in) at (1,0);
 			\coordinate (-closed) at (-1,1);
 			\coordinate (-opened) at (-1,-1);
 			
 			\draw [line cap=rect] (-1,1) -- (-1,0.5);
 			\draw [line cap=rect] (-1,-1) -- (-1,-0.5);
 			\draw [line cap=round] (1, 0) -- ($(1, 0)!2/3!(-1.3,0.8)$);
 			\draw [line cap=rect] ($(1, 0)!1/3!(-1.3,0.8)$) -- (-1.3,0.8);
 			%			\draw [line cap=round] (1, 0) -- ($(1, 0)!2/3!(-1.3,-0.8)$);
 			%			\draw [line cap=rect] ($(1, 0)!1/3!(-1.3,-0.8)$) -- (-1.3,-0.8);
 	}}
}

\begin{tikzpicture}
	\node[block, align=center] (wpe) {Dereverberate};
	\node[block, align=center, anchor=west] (bf) at ($(wpe.east) + (4em,0)$) {Beamforming};
	\node[block, align=center, anchor=north] (bfest) at ($(bf.south) + (0, -1.4em)$) {Estimate $\vect{q}$};
	\node[block, align=center, anchor=north] (WPEfilter) at ($(wpe.south) + (0, -1.4em)$) {Estimate $\Gstacked$};
	\node[block, align=center, anchor=north] (psd) at ($(WPEfilter.south) + (0, -1.7em)$) {Estimate $\lambda_t$};
	
% 	\pic [rotate=180, yscale=1] (switch) at ($(psd.south) + (0, -2em)$) {pic switch={1.5em times 1.5em}};
	\node[block, align=center, anchor=north] (atf) at ($(bfest.south) + (0em, -1.7em)$) {RTF estimate};
	
	\draw[arrow] ($(atf.south) + (0em, -1em)$) node[below] (noiseonly) {Noise-only periods} -- (atf);
	
	\coordinate (bottom) at ($(noiseonly.south) + (0, 0em)$);
	
	\pic [rotate=0, yscale=1] (switchInit) at ($(psd.west) + (-0.5,0)$) {pic switch={1.5em times 1.5em}};
	
	\coordinate (y) at ($(wpe.west) + (-5em, 0)$);
	\coordinate (d) at ($(wpe.east)!1/2!(bf.west)$);
	
	\draw[arrow] (y) node[above right]{$\y_t$} -- (wpe);
	\draw[arrow] (wpe) -- (bf);
	\node[above] (dtext) at (d) {$\dhat_t$};
	\draw[arrow] (bf.east) -- +(2.5em, 0) node[above left]{$z_t$};
	
	\draw[] ($(bf.east) + (1em, 0)$) node[branch]{} |- (bottom) -| (switchInit-south west);
% 	\draw[] (switch-east) -| (switchInit-south west);
	\draw[] (y-|switchInit-closed) node[branch]{} -- (switchInit-closed);
	\draw[arrow] (switchInit-in) -- (psd);
	
	\node[left] (inittext) at (switchInit-closed) {\tiny init};
%	\node[right] at (switch-west) {\tiny option};
% 	\node[above right] at (switch-closed) {\tiny factorized};
% 	\node[above right] at (switch-opened) {\tiny cascaded};
	
	\draw[arrow] (psd.north) node[above right]{$\hat\lambda_t$} -- (WPEfilter);
	\draw[arrow] (WPEfilter) -- node[right]{$\Gstacked$} (wpe);
	\draw[arrow] (switchInit-closed |- WPEfilter) node[branch]{} -- (WPEfilter);

	\draw[arrow] (d) node[branch]{} -| (atf-|d) -- (atf);	
% 	\draw[] (atf-|d) node[branch]{} |- (switch-opened);	
	
	\coordinate (tmp) at ($(bfest.north west)!1/3!(bfest.south west)$);
	\draw[arrow] (d |- tmp) node[branch]{} -- (tmp);
	\coordinate (tmp) at ($(bfest.north west)!2/3!(bfest.south west)$);
	\coordinate (tmp2) at ($(wpe.east)!1/3!(bf.west)$);
	\draw[arrow] (psd) -| (tmp2 |- tmp) -- (tmp);
	
	\draw[arrow] (atf.north) node[above right]{$\tilde{\vect{v}}$} -- (bfest);
	\draw[arrow] (bfest) -- node[right]{$\vect{q}$} (bf);
	
	\tikzstyle{boundingBox}=[inner sep=0.4em, draw, dashed, black!70, rounded corners=0.3em, shift={(0,0em)}]
	
    % 	(switchInit-opened)(switchInit-closed)(d)(dtext.center)(inittext.center)
	\node[boundingBox, fit=(wpe)(WPEfilter)(psd)] (wpeBox) {};
	\node[text=black!70, above, anchor=south] at (wpeBox.north) {WPE};
	
	\node[boundingBox, fit=(bf)(bfest)] (wMPDRBox) {};
	\node[text=black!70, above, anchor=south] at (wMPDRBox.north) {wMPDR};
	
\end{tikzpicture}
	\caption{Proposed factorisation of \gls{WPD} in \gls{WPE} and \gls{wMPDR} with \gls{RTF} estimation and power estimation (joint optimization).}
	\label{fig:WPEwMPDR}
\end{figure}
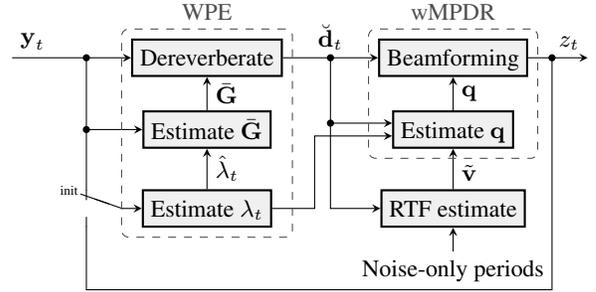

\begin{figure}
    \centering
    \begin{subfigure}{0.49\columnwidth}\centering
\begin{tikzpicture}
    \node[block] (wpe) {WPE};
    \node[block, anchor=west] (bf) at ($(wpe.east) + (2em,0)$) {BF};
    \coordinate (up) at ($(wpe.north)!1/3!(wpe.south)$);
    \coordinate (down) at ($(wpe.north)!2/3!(wpe.south)$);
    \coordinate (bottom) at ($(wpe.south) + (0, -0.6em)$);
    \draw[arrow] ($(wpe.west |- up) + (-3em, 0)$) node[above right]{$\y_t$} -- (wpe.west |- up);
    \draw[arrow] (wpe.east |- up) -- (bf.west |- up);
    \draw[arrow] (bf.east |- up) -- +(2em, 0) node[above left]{$z_t$};
    \draw[arrow] (wpe.east |- down) -| +(0.8em, 0) |- ($(wpe.west |- bottom) + (-0.8em, 0)$) node[above left] {$\lambda_t$} |- (wpe.west |- down);
    \draw[arrow] (bf.east |- down) -| ($(bf.east |- bottom) + (0.8em, 0)$) node[above right] {$\lambda_t$} -- ($(bf.west |- bottom) + (-0.8em, 0)$) |- (bf.west |- down);
\end{tikzpicture}
\subcaption{Separate optimization}
\label{fig:cascaded}
\end{subfigure}
\begin{subfigure}{0.49\columnwidth}\centering
\begin{tikzpicture}
    \node[block] (wpe) {WPE};
    \node[block, anchor=west] (bf) at ($(wpe.east) + (2em,0)$) {BF};
    \coordinate (up) at ($(wpe.north)!1/3!(wpe.south)$);
    \coordinate (down) at ($(wpe.north)!2/3!(wpe.south)$);
    \coordinate (bottom) at ($(wpe.south) + (0, -0.6em)$);
    \draw[arrow] ($(wpe.west |- up) + (-3em, 0)$) node[above right]{$\y_t$} -- (wpe.west |- up);
    \draw[arrow] (wpe.east |- up) -- (bf.west |- up);
    \draw[arrow] (bf.east |- up) -- +(2em, 0) node[above left]{$z_t$};
    \draw[arrow] (bf.east |- down) -| +(0.8em, 0) |- ($(wpe.west |- bottom) + (-0.8em, 0)$) node[above left] {$\lambda_t$} |- (wpe.west |- down);
\end{tikzpicture}
\subcaption{Joint optimization}
\label{fig:integrated}
\end{subfigure}
    \caption{Separate and joint optimization schemes.
    %In a cascaded combination of WPE and beamforming (BF) the power estimation for WPE is done on the WPE output, while in the integration %approach the BF output is used.
    %\inred{Feel free to modify or delete this figure.}
    }
    \label{fig:iterations}
\end{figure}
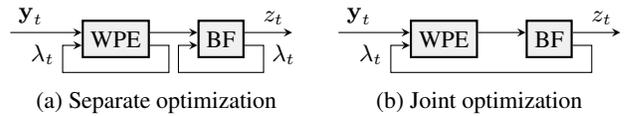

In this section, we experimentally confirm the equivalence of the unified and factorized solutions, and present detailed analysis of the factorized solution.

\subsection{Equivalence experiment}
We first show the equivalence of the unified and factorized solution experimentally by applying both to a CHiME3 utterance \cite{Barker2015CHiME3}.
%Using the estimation scheme visualized in \cref{fig:WPEwMPDR} for one iteration, 
% Assuming that the \gls{RTF} is estimated with the help of oracle energy ratio masks (Wiener like \cite{Erdogan2015masks}) as it is done, and that the \gls{PSD} of the source, $\lambda_t$, is determined as the power of observation, we obtain an estimate for the factorized solution $z_t^{\mathrm{factorized}}$ and that for the unified solution $z_t^{\mathrm{unified}}$ for each frequency.
The \gls{RTF} is estimated with the help of oracle energy ratio masks (Wiener like \cite{Erdogan2015masks}), and the \gls{PSD} of the source, $\lambda_t$, is determined as the power of the observation.
Using these values we obtain an estimate for the factorized solution $z_t^{\mathrm{factorized}}$ and for the unified solution $z_t^{\mathrm{unified}}$ for each frequency.
%For the unified solution  we reused the \gls{RTF} and \gls{PSD} estimation.
Then, to verify the equivalence, we tested that the following inequality holds for every time-frequency point in the \gls{STFT}:
\begin{align}
    \frac{\left\lVert z_t^{\mathrm{factorized}} - z_t^{\mathrm{unified}} \right\rVert}{
        % \frac{1}{T}\sum_{\tilde{t}} z_{\tilde{t}}^{\mathrm{factorized}}
        % \operatornamewithlimits{max}\limits_{\tilde{t}}{
            \left\lVert z_{\tilde{t}}^{\mathrm{factorized}}\right\rVert
        % }
    } \leq 10^{-9}
\end{align}
A maximum relative difference of $10^{-9}$ is reasonable for double-precision floating-point values, where different mathematical calculations are used (e.g. in both solutions a linear system of equations has to be solved, but in the unified solution there are more linear equations: \eqref{eq:wpdfilter} vs \eqref{eq:Gpast}).

\subsection{Experimental analysis of proposed factorization}
In the following, we present an experimental analysis of the proposed factorization (WPE+wMPDR) by comparing it with the conventional cascade configuration (WPE+MPDR) and various beamforming configurations, including \gls{MPDR}, MVDR, and \gls{wMPDR}.

\subsubsection{Dataset, evaluation metrics, and analysis conditions}
\begin{table}[!t]
%% increase table row spacing, adjust to taste
\renewcommand{\arraystretch}{1.0}
\caption{WERs (\%) of enhanced speech obtained after 1st iteration. Boldface indicates the best score for each condition. }\label{tbl:aq}
\label{table_example}
\centering
\begin{tabular}{ccc}%\hline
\toprule
& \multicolumn{2}{c}{Real eval set}\\ %\hline
\cmidrule{2-3}
& CHiME3 & REVERB\\ %\hline
\midrule
Obs & 17.83 & 18.61\\ %\hline
MPDR & 7.47 & 13.14 \\
MVDR & 7.50 & 12.87 \\
wMPDR & {\bf 6.99} & {\bf 12.65} \\ %\hline
\cmidrule{1-3}
WPE & 13.95 & 13.24 \\ %\hline
\cmidrule{1-3}
WPE+MPDR (joint opt.) & 7.55 & 10.06 \\
WPE+wMPDR (joint opt.) & {\bf 7.07} & {\bf 9.52} \\
%WPD & {\bf 7.01} & 9.56 \\
% \hline
\bottomrule
\end{tabular}
\end{table}

For the analysis, we used the REVERB Challenge dataset (REVERB) \cite{REVERB2016} and the CHiME3 challenge dataset (CHiME3) \cite{Barker2015CHiME3}. 
Each utterance in REVERB was recorded in reverberant environments with a little stationary additive noise, while that in CHiME3 was recorded in public areas with relatively high level non-stationary ambient noise and a little reverberation.
Separate optimization of WPE and MVDR/MPDR has been shown to be very effective as frontend of ASR for both dataset \cite{REVERB2016,Barker2015CHiME3}.
%\inred{CB: Do we have something to cite this statement? Did the WPE iteration include the MVDR for the power estimation?}
%\inblue{TN: NTT Reverb/Chime3 systems used wpe followed by mvdr/mpdr. As you mention, the WPE iteration did not include MVDR/MPDR for the power estimation, I replaced WPE+MVDR with WPE followed by MVDR/MPDR.}

For the performance evaluation, we used baseline ASR systems recently developed using Kaldi \cite{Povey2011Kaldi}, respectively, for REVERB and CHiME3. They are fairly competitive systems composed of a TDNN acoustic model trained using lattice-free MMI, online i-vector extraction, and a trigram language model.

A Hann window was used for a short-time analysis with the sampling frequency being 16 kHz. $M=8$ and $M=6$ microphones were used, respectively, for REVERB and CHiME3.
The prediction delay was set at $b=4$, and the length of the prediction filter was set at $L_w=12, 10$, and $6$, respectively, for frequency ranges of $0$ to $0.8$ kHz, $0.8$ to $1.5$ kHz, $1.5$ to $8$ kHz, for REVERB, and $L_w=7$ for CHiME3.
%\inred{CB: Do you use only $4$ values of the history in WPE for CHiME3? 
%I expected a higher number (I assumed the stft size to be 512 and the shift to be 128).
%According to http://www.kecl.ntt.co.jp/icl/signal/wpe/config.html your MATLAB code takes $L_w-b+1$ as argument.}

\subsubsection{Estimation of power spectral density and RTF}
Figure 1 illustrates the overall processing flow of the estimation, where we jointly estimate the \gls{PSD}, $\lambda_t$, and the \gls{RTF}, $\tilde{\mathbf{v}}$. 
%
%We have two alternatives to estimate the convolutional beamformer (unified and factorized), but we assumed the time-variant \gls{PSD} $\lambda_t$ is known.
The \gls{PSD} $\lambda_t$ is estimated with the same \gls{ML} objective, but since no closed-form solution is known, the \gls{PSD} $\lambda_t$ and the convolutional beamformer are estimated alternatingly based on iterative optimization \cite{nakatani2019maximum}.
% The relevant part of the likelihood from \cite{nakatani2019maximum} is:
% \begin{align}
% 	\mathcal{L}(\wstacked, \lambda_t) &= -\frac{1}{T} \sum_{t=1}^{T} \left(\frac{\left|z_t\right|^2}{\lambda_t} + \ln(\lambda_t)\right)
% \end{align}
Maximizing \eqref{eq:likelihood} w.r.t. $\lambda_t$ will yield
\begin{align}
    \lambda_t
    = \left|z_t\right|^2
    = \left|\wstacked\H \ystacked_t\right|^2 .
\end{align}
This parameter estimation is referred to as joint optimization scheme shown in \cref{fig:integrated}.
%, where \gls{WPE} and beamforming is optimized depending on each other.
In the experiments, we also test a separate optimization scheme shown in (\cref{fig:cascaded}), where the \gls{WPE} and beamforming are optimized separately and the \gls{PSD} is estimated using iterative optimization of respective processing blocks.
%\begin{align}
%    \lambda_t
%    = \left\lVert\dhat_t\right\rVert^2
%    = \left\lVert\Gstacked\H \ystacked_t\right\rVert^2.
%\end{align}

For the estimation of the \gls{RTF} $\tilde{\mathbf{v}}$, we used a method based on eigenvalue decomposition with noise covariance whitening \cite{ito17icassp,Markovich2015performance}, and apply it to the output of \gls{WPE} dereverberation, to reduce the effect of reverberation and noise from the estimation. For estimation of noise spatial covariance matrices, we assumed that each utterance had noise-only periods of 225 ms and 75 ms, respectively, at its beginning and ending parts, for REVERB, and we used noise masks estimated by a BLSTM network \cite{Heymann2017Beamnet} for CHiME3.
%For WPE and WPE+MPDR, we updated all the parameters in each iteration.  

\subsubsection{Evaluation results}
Table~\ref{tbl:aq} summarizes the WERs of the observed signals (Obs) and the enhanced signals obtained after the first estimation iteration. Here, we used the joint optimization scheme for WPE+wMPDR.
In the table, WPE+wMPDR was the best among all the methods.
%This proves that WPD can be factorized into WPE and wMPDR without loss of optimality.
When we compare wMPDR with MPDR/MVDR, and compare WPE+wMPDR with WPE+MPDR, wMPDR and WPE+wMPDR consistently outperformed MPDR/MVDR and WPE+MPDR, respectively.
This shows that the superiority of the convolutional beamformer is surely derived from the superiority of wMPDR embedded into it.

\begin{figure}[!t]
  \centering
  \includegraphics[width=3in]{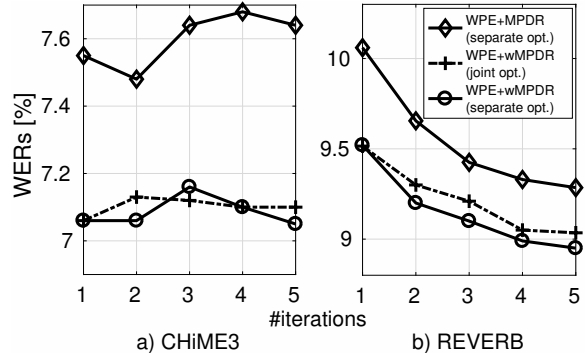}
  \caption{WERs (\%) obtained with \# of estimation iterations by WPE+MPDR and WPE+wMPDR with joint and separate optimization schemes}\label{fig:pcurve}
\end{figure}

Figure ~\ref{fig:pcurve} shows the WERs obtained by iterative estimation. In addition to the joint optimization scheme for the estimation of $\lambda_t$, we used the separate optimization scheme. 
With the separate optimization, a specified number of iterations is first performed by WPE and then performed by beamforming.
We test this configuration as a simpler alternative of WPE+wMPDR. As shown in the figure, with both joint and separate optimization schemes, the proposed factorization, WPE+wMPDR, achieved almost the same WERs, and they are consistently better than the conventional cascade configuration, WPE+MPDR, with the separate optimization.

\section{Conclusions}
\label{sec:conclusions}
In this contribution we factorized the \gls{WPD} convolutional beamformer in \gls{WPE} dereverberation and \gls{wMPDR} beamforming and displayed practical advantages.
The equivalence is verified mathematically and numerically.
% The factorization has practical advantages, because the 
In experiments on real data we showed that the strength of the \gls{WPD} convolutional beamformer has its origin in the \gls{wMPDR} beamformer.
% In a comparison if \gls{WPE} and \gls{wMPDR} should be separate  
A comparison of a simple separate optimization with a joint optimization of \gls{WPE} and \gls{wMPDR} yielded similar \glspl{WER}.

%%%%%%%%%%%%%%%%%%%%%%%%%%%%%%%%%%%%%%%%%%%%%%%%%%%%%%%%
\section{Appendix: Unified versus factorized solution}
\label{sec:unified_factorized}
%%%%%%%%%%%%%%%%%%%%%%%%%%%%%%%%%%%%%%%%%%%%%%%%%%%%%%%%

We now show  that the two solutions, Sec.~\ref{sec:unified} and Sec.~\ref{sec:cascaded} , are identical. Starting with the factorized solution, note first that $\Rd= \Gstacked\H\Rstacked\Gstacked$ can be expressed as
\begin{align}
	\label{eq:Rd_via_Ry}
		\Rd = \left(\R-\P\H\Rpast\inv\P\right)
\end{align}
using \eqref{eq:Gstacked}, \eqref{eq:Rstacked} and \eqref{eq:Gpast}. Employing this in \eqref{eq:factorized_1} we can express the convolutional beamformer coefficients as
\begin{align}
  \label{eq:wstacked_extended}
	\wstacked &= \Gstacked\bfonly \nonumber\\
	&= \begin{bmatrix} 	\vect{I} \\ -\Rpast\inv\P \end{bmatrix}
	\frac{\left(\R-\P\H\Rpast\inv\P\right)\inv\rtf}{\rtf\H\left(\R-\P\H\Rpast\inv\P\right)\inv\rtf}
\end{align}
where we expressed $\Gstacked$ using \eqref{eq:Gstacked} and \eqref{eq:Gpast}, and $\bfonly$ using \eqref{eq:mvdr_like}.

On the other hand, we take the unified solution \eqref{eq:wpdfilter} and plug in the definition of $\Rstacked$, eq.~\eqref{eq:Rstacked}. Employing the the $(2 \times 2)$ block matrix inversion rule, we exactly obtain \eqref{eq:wstacked_extended}. Thus, the solutions \eqref{eq:wpdfilter} and \eqref{eq:wstacked_extended} are identical!

\vfill\pagebreak

%List and number all bibliographical references at the end of the
%paper. The references can be numbered in alphabetic order or in
%order of appearance in the document. When referring to them in
%the text, type the corresponding reference number in square
%brackets as shown at the end of this sentence \cite{C2}. An
%additional final page (the fifth page, in most cases) is
%allowed, but must contain only references to the prior
%literature.

\balance
\bibliographystyle{IEEEbib}
\bibliography{IEEEabrv,strings,refs}

\begin{thebibliography}{10}

\bibitem{LiDeHaGo2015}
J.~Li, L.~Deng, R.~Haeb-Umbach, and Y.~Gong,
\newblock {\em Robust Automatic Speech Recognition; ch. 9: Reverberant Speech
  Recognition},
\newblock Elsevier, Oct 2015.

\bibitem{Lebart2001AA}
K.~Lebart, J.~M. Boucher, and P.~Denbigh,
\newblock ``A new method based on spectral subtraction for speech
  dereverberation,''
\newblock {\em Acta Acoustica}, vol. 87, no. 3, pp. 359–366, 2001.

\bibitem{Schwarts2015TASLP}
O.~Schwartz, S.~Gannot, and E.~A.~P. Habets,
\newblock ``Multi-microphone speech dereverberation and noise reduction using
  relative early transfer function,''
\newblock {\em {IEEE/ACM} Trans. Audio, Speech, Lang. Process.}, vol. 23, no.
  2, pp. 240--251, 2015.

\bibitem{Nakatani2010TASLP}
T.~Nakatani, T.~Yoshioka, K.~Kinoshita, M.~Miyoshi, and B.-H. Juang,
\newblock ``Speech dereverberation based on variance-normalized delayed linear
  prediction,''
\newblock {\em {IEEE/ACM} Trans. Audio, Speech, Lang. Process.}, vol. 18, no.
  7, pp. 1717--1731, 2010.

\bibitem{Yoshioka2012GeneralWPE}
T.~Yoshioka and T.~Nakatani,
\newblock ``Generalization of multi-channel linear prediction methods for blind
  {MIMO} impulse response shortening,''
\newblock {\em {IEEE/ACM} Trans. Audio, Speech, Lang. Process.}, 2012.

\bibitem{REVERB2016}
K.~Kinoshita, M.~Delcroix, S.~Gannot, E.~Habets, R.~Haeb-Umbach, W.~Kellermann,
  V.~Leutnant, R.~Maas, T.~Nakatani, B.~Raj, A.~Sehr, and T.~Yoshioka,
\newblock ``A summary of the {REVERB} challenge: state-of-the-art and remaining
  challenges in reverberant speech processing research,''
\newblock {\em EURASIP Journal on Advances in Signal Processing}, 2016.

\bibitem{VanVeen1988Beamforming}
B.~D. {Van Veen} and K.~M. {Buckley},
\newblock ``Beamforming: A versatile approach to spatial filtering,''
\newblock {\em IEEE ASSP Magazine}, vol. 5, no. 2, pp. 4--24, April 1988.

\bibitem{Erdogan2015masks}
H.~Erdogan, J.~R. Hershey, S.~Watanabe, and J.~Le~Roux,
\newblock ``Phase-sensitive and recognition-boosted speech separation using
  deep recurrent neural networks,''
\newblock in {\em 2015 IEEE International Conference on Acoustics, Speech and
  Signal Processing (ICASSP)}. IEEE, 2015, pp. 708--712.

\bibitem{Barker2015CHiME3}
J.~Barker, R.~Marxer, E.~Vincent, and S.~Watanabe,
\newblock ``The third {‘CHiME’} speech separation and recognition
  challenge: Dataset, task and baselines,''
\newblock in {\em 2015 IEEE Workshop on Automatic Speech Recognition and
  Understanding (ASRU)}. IEEE, 2015, pp. 504--511.

\bibitem{Kanda2019}
N.~Kanda, C.~Boeddeker, J.~Heitkaemper, Y.~Fujita, S.~Horiguchi, K.~Nagamatsu,
  and R.~Haeb-Umbach,
\newblock ``Guided source separation meets a strong {ASR} backend:
  {H}itachi/{P}aderborn {U}niversity joint investigation for dinner party
  {ASR},''
\newblock in {\em Proc. \mbox{Interspeech}}, Sep 2019.

\bibitem{Braun2018TASLP}
S.~Braun and E.~A.~P. Habets,
\newblock ``Linear prediction-based online dereverberation and noise reduction
  using alternating {Kalman} filter,''
\newblock {\em {IEEE/ACM} Trans. Audio, Speech, Lang. Process.}, vol. 28, no.
  6, pp. 1119--1129, 2018.

\bibitem{Drude2018integration}
L.~Drude, C.~Boeddeker, J.~Heymann, R.~Haeb-Umbach, K.~Kinoshita, M.~Delcroix,
  and T.~Nakatani,
\newblock ``Integrating neural network based beamforming and weighted
  prediction error dereverberation.,''
\newblock in {\em Interspeech}, 2018, pp. 3043--3047.

\bibitem{Delcroix2015ESP}
M.~Delcroix, T.~Yoshioka, A.~Ogawa, Y.~Kubo, M.~Fujimoto, I.~Nobutaka,
  K.~Kinoshita, M.~Espi, T.~Hori, and T.~Nakatani,
\newblock ``Strategies for distant speech recognition in reverberant
  environments,''
\newblock {\em EURASIP Journal on Advances in Signal Processing}, 2015.

\bibitem{Nakatani2019AUC}
T.~Nakatani and K.~Kinoshita,
\newblock ``A unified convolutional beamformer for simultaneous denoising and
  dereverberation,''
\newblock {\em IEEE Signal Processing Letters}, vol. 26, pp. 903--907, April
  2019.

\bibitem{nakatani2019maximum}
T.~Nakatani and K.~Kinoshita,
\newblock ``Maximum likelihood convolutional beamformer for simultaneous
  denoising and dereverberation,''
\newblock in {\em Proc. EUSIPCO}, A Coruna, Spain, Sep. 2019.

\bibitem{Petersen2008MCB}
K.~B. Petersen, M.~S. Pedersen, et~al.,
\newblock ``The matrix cookbook,''
\newblock {\em Technical University of Denmark}, vol. 7, no. 15, pp. 510, 2008.

\bibitem{Drude2018narawpe}
L.~Drude, J.~Heymann, C.~Boeddeker, and R.~Haeb-Umbach,
\newblock ``{NARA-WPE}: A {Python} package for weighted prediction error
  dereverberation in {Numpy} and {Tensorflow} for online and offline
  processing,''
\newblock in {\em Speech Communication; 13th ITG-Symposium}. VDE, 2018, pp.
  1--5.

\bibitem{Boeddeker2018MA}
C.~Boeddeker, L.~Drude, and R.~Haeb-Umbach,
\newblock ``Optimization of multi-channel dereverberation techniques for noisy
  reverberant speech recognition,''
\newblock M.S. thesis, Paderborn University, 2018.

\bibitem{Cho2019MLDR}
B.~J. {Cho}, J.~{Lee}, and H.~{Park},
\newblock ``A beamforming algorithm based on maximum likelihood of a complex
  {Gaussian} distribution with time-varying variances for robust speech
  recognition,''
\newblock {\em IEEE Signal Processing Letters}, vol. 26, no. 9, pp. 1398--1402,
  Sep. 2019.

\bibitem{Povey2011Kaldi}
D.~Povey, A.~Ghoshal, G.~Boulianne, L.~Burget, O.~Glembek, N.~Goel,
  M.~Hannemann, P.~Motlicek, Y.~Qian, P.~Schwarz, J.~Silovsky, G.~Stemmer, and
  K.~Vesely,
\newblock ``The {Kaldi} speech recognition toolkit,''
\newblock in {\em IEEE 2011 Workshop on Automatic Speech Recognition and
  Understanding}. Dec. 2011, IEEE Signal Processing Society,
\newblock IEEE Catalog No.: CFP11SRW-USB.

\bibitem{ito17icassp}
N.~Ito, S.~Araki, M.~Delcroix, and T.~Nakatani,
\newblock ``Probabilistic spatial dictionary based online adaptive beamforming
  for meeting recognition in noisy and reverberant environments,''
\newblock in {\em 2017 IEEE International Conference on Acoustics, Speech and
  Signal Processing (ICASSP)}.

\bibitem{Markovich2015performance}
S.~Markovich-Golan and S.~Gannot,
\newblock ``Performance analysis of the covariance subtraction method for
  relative transfer function estimation and comparison to the covariance
  whitening method,''
\newblock in {\em 2015 IEEE International Conference on Acoustics, Speech and
  Signal Processing (ICASSP)}. IEEE, 2015, pp. 544--548.

\bibitem{Heymann2017Beamnet}
J.~Heymann, L.~Drude, C.~Boeddeker, P.~Hanebrink, and R.~Haeb-Umbach,
\newblock ``Beamnet: End-to-end training of a beamformer-supported
  multi-channel {ASR} system,''
\newblock in {\em 2017 IEEE International Conference on Acoustics, Speech and
  Signal Processing (ICASSP)}. IEEE, 2017, pp. 5325--5329.

\end{thebibliography}
\end{document}